\documentclass[12pt]{article}
\usepackage{amsfonts,amsmath}
\usepackage{amssymb}

\makeatletter \@addtoreset{equation}{section} \makeatother

\newcommand{\dr}{{\rm d}}
{\vspace{3mm} }

\def\al{\alpha}

\def\*{\star}

\def\E2{\mathbf{E}}

\newcommand{\be}{\begin{equation}}
\newcommand{\ee}{\end{equation}}
\newcommand{\bee}{\begin{eqnarray}}
\newcommand{\beee}{\begin{array}}
\newcommand{\eee}{\end{eqnarray}}
\newcommand{\eeee}{\end{array}}
\newcommand{\gb}{\beta}
\newcommand{\gga}{\gamma}

\newcommand{\W}{{\cal W}}
\newcommand{\Lc}{{\cal L}}

\newcommand{\gd}{\delta}
\newcommand{\gl}{\lambda}
\newcommand{\gk}{\varkappa}
\newcommand{\gep}{\epsilon}

\newcommand{\go}{\omega}

\newcommand{\dal}{\dot \alpha}
\newcommand{\dgb}{\dot \beta}
\newcommand{\dgga}{\dot \gamma}

\newcommand{\p}{\partial}

\newcommand{\ff}{\frac}

\begin{document}
\begin{flushright}
FIAN/TD/2015-19\\
\end{flushright}

\vspace{0.5cm}
\begin{center}
{\large\bf Charges in nonlinear higher-spin theory}

\vspace{1 cm}

\textbf{V.E.~Didenko,}\textsc{$^{1}$}\textbf{ N.G.~Misuna}\textsc{$^{1,2}$}\textbf{
and M.A.~Vasiliev}\textsc{$^{1}$}\\
 \vspace{0.5cm}
 \textsc{$^{1}$} \textit{I.E. Tamm Department of Theoretical Physics,
Lebedev Physical Institute,}\\
 \textit{ Leninsky prospect 53, 119991, Moscow, Russia}\\

\par\end{center}

\begin{center}
\textsc{$^{2}$}\textit{Moscow Institute of Physics and Technology,}\\
 \textit{ Institutsky lane 9, 141700, Dolgoprudny, Moscow region,
Russia}\\

\par\end{center}

\begin{center}
\vspace{0.6cm}
 didenko@lpi.ru, misuna@phystech.edu, vasiliev@lpi.ru \\

\par\end{center}

\vspace{0.4cm}

\begin{abstract}
\noindent Nonlinear higher-spin equations in four dimensions admit
a closed two-form that defines a gauge-invariant global charge as
an integral over a two-dimensional cycle. In this paper we argue
that this charge gives rise to partitions depending on various
lower- and higher-spin chemical potentials identified with modules
of topological fields in the theory. The vacuum contribution to
the partition is calculated to the first nontrivial order for a
solution to higher-spin equations that generalizes $AdS_4$ Kerr
black hole of General Relativity. The resulting partition is
non-zero being in parametric  agreement with the ADM-like behavior
of a rotating source. The linear response of chemical potentials
to the partition function is also extracted. The explicit unfolded
form of $4d$ GR black holes is given.
An explicit formula relating
 asymptotic higher-spin charges expressed in terms of the generalized
 higher-spin Weyl tensor with those expressed in terms of Fronsdal fields is obtained.
\end{abstract}

\section{Introduction}
Higher-spin (HS) gauge theories \cite{Vasiliev:1999ba, RevD}
have recently attracted much of attention in the
context of $AdS/CFT$ correspondence (see e.g. \cite{GYRev}
 and references therein). The conjecture of Klebanov and
Polyakov \cite{KP, SS} associates HS theories with
vectorial models on the flat boundary and provides an example of
the weak-weak type duality that can be verifiable in
practice. The main obstacle that however prevents one from solid
tests of the correspondence is the absence of the full nonlinear
extension of the free Fronsdal
action principle on
the HS side \cite{fronsdal} (see however \cite{BSaction}). The available tests
favoring to Klebanov-Polyakov conjecture were completed either
at the level of equations of motion \cite{GY1, GY2} or by implying
symmetry arguments \cite{MZh1, MZh2}. Recently an on-shell
invariant functional has been proposed in \cite{Vasiliev:1504},
conjectured to produce the generating functional for boundary
correlation functions. As such, it may provide a proper substitute
of  action
principle for the $AdS/CFT$ HS problem. In $d=4$ the invariant
functional is a closed space-time four-form which arises upon an
extension of the original HS equations \cite{more} by adding certain
auxiliary extra fields.

Apart from the four-form anticipated to be relevant to the $AdS/CFT$ dictionary,
there is also a closed two-form defined on the equations of
motion. Being integrated over a two-dimensional cycle it provides
a conserved charge on solutions of HS equations. Moreover, this construction
allows a straightforward inclusion of HS topological fields introduced in
\cite{more}, which can be identified with various chemical potentials.

In this paper we analyze the topological moduli contribution at first order in
the linearized approximation. We show how the obtained formulae
reproduce asymptotic charges associated with moduli parameters of
arbitrary rank. In general this results in some higher-derivative
expression for a closed two-form which is hard to derive using
standard General Relativity (GR) methods. We also consider the example of gravity in
which case one reproduces asymptotic charge of a kind considered
in \cite{Ashtekar}.

To explain the idea of our approach we first review the generating construction of GR BHs on
$AdS$ along with their HS generalization in terms of a single
$AdS$ global symmetry parameter  \cite{DMV}. We put emphasis on
particular cases of most physical relevance that are Kerr,
Schwarzschild, planar and topological BHs. We show that the charge
calculated this way for the Kerr case parametrically agrees with
the standard asymptotic ADM-like behavior.

The paper is organized as follows. In section 2 we review $4d$
HS equations and define the invariant two-form functional. In
section 3 we introduce the notion of asymptotic symmetries in HS
theory and introduce charges and partitions with the emphasis on
HS topological sector. Then in section 3.1 the simplest vacuum
partition is further derived at free level. In section 3.2 the
topological contribution is found to the lowest order and the
examples of spin two and spin four are considered. In section 4 we review how
$AdS$ BHs and their HS cousins originate from $AdS$ global
symmetry parameter and consider various BH examples in detail.
Finally in section 4.3 the charge for the Kerr case is explicitly
computed.

\section{Structure of HS equations}

HS equations in four dimensions have the following standard form \cite{more}
\begin{align}
&\dr W+W*W=0\,,\label{HS1}\\
&\dr S+[W,S]_*=0\,,\label{HS2}\\
&\dr B+[W,B]_*=0\,,\label{HS3}\\
&S*S=-i\theta_{\al}\wedge \theta^{\al}(1+F_*(B)*k\gk)-
i\bar\theta_{\dal}\wedge \bar\theta^{\dal}(1+\bar F_*(B)*\bar k\bar \gk)\,,\label{HS4}\\
&[S,B]_*=0\,.\label{HS5}
\end{align}
The conventions are  as follows. $W$, $S$ and $B$ in
\eqref{HS1}-\eqref{HS5} denote space-time dependent generating
functions of commuting twistor-like variables $Y_{A}=(y_{\al},
\bar{y}_{\dal})$ and $Z_{A}=(z_{\al} ,\bar{z}_{\dal})$, where
spinor indices $\al, \gb,...$ range two values. The associative
star-product operation acts on functions of $(Y,Z)$-space
\be\label{star}
(f*g)(Y,Z)=\ff{1}{(2\pi)^4}\int dU dV f(Y+U, Z+U)g(Y+V,
Z-V)e^{iU_A V^A}\,,
\ee
where $U_AV^A:=U_A V_B \gep^{AB}$ with some $sp(4)$-invariant
form $\gep_{AB}=-\gep_{BA}$. Indices are raised and lowered with
the aid of $\gep_{AB}$ as follows, $X^{A}=\gep^{AB}X_{B}$ and
$X_{A}=\gep_{BA}X^{B}$. The integration measure is chosen in such
a way that
\be
1*F(Y,Z)=F(Y,Z)*1=F(Y,Z)\,.
\ee
The star product induces the following commutation relations
\be
[Y_A ,Y_B]_*=-[Z_{A}, Z_{B}]_*=2i\gep_{AB}\,,\qquad [Y_A,
Z_B]_*=0\,.
\ee
Important elements of the star-product algebra entering
\eqref{HS4} are the inner Klein operators $\gk$ and $\bar\gk$
\be
\gk=\exp iz_{\al}y^{\al}\,,\qquad \bar\gk=\exp
i\bar{z}_{\dal}\bar y^{\dal}\,.
\ee
Using \eqref{star} it is easy to check  their characteristic properties
\be\label{Klpr}
\{\gk, y_{\al}\}_*=\{\gk, z_{\al}\}_*=0\,,\qquad \gk*\gk=1\,,
\ee
analogously in the antiholomorphic sector for $\bar{\gk}$.

There are also extra Clifford variables $K=(k, \bar k)$ which enter
system \eqref{HS1}-\eqref{HS5}. $k$ anticommutes with the holomorphic
variables $y_{\al}, z_{\al}, \theta_{\al}$ and commutes with every
antiholomorhic one, while $\bar k$ does the opposite.
In addition
\be\label{k^2}
k^2=\bar k^2=1\,.
\ee
Note that along with \eqref{Klpr} this  makes elements $k\gk$ and $\bar k
\bar\gk$ central in the star-product algebra. That
\be
\{\theta, k\}=\{\bar\theta, \bar k\}=0
\ee
does not allow one to realize $k$ and $\bar k$ in terms of
star-product variables.

A space-time one-form $W(Y,Z; K|x)=W_{\mu}dx^{\mu}$ parameterizes
on-shell HS potentials and their derivatives. The 0-form
$B(Y,Z; K|x)$ encodes spin $s=0$, $s=1/2$ matter fields and all HS
curvatures. The $S(Y,Z; K|x)=S_A \theta^A$ field, where
$\theta_A=(\theta_{\al}, \bar\theta_{\dal})$, representing a one-form with
respect to spinorial differentials, is purely auxiliary on-shell carrying
no local degrees of freedom. All differentials anticommute
\be
\{dx_\mu, dx_\nu\}=\{dx_\mu, \theta_A\}=\{\theta_A,
\theta_B\}=0\,.
\ee

By virtue of \eqref{k^2} the dependence of $W$, $B$ and $S$ on $K$
is at most bilinear,
\be
W=\sum_{i,j=0,1}W_{i,j}k^i\bar k^{j}\,,\qquad
B=\sum_{i,j=0,1}B_{i,j}k^i\bar k^{j}\,.
\ee
Such dependence is known to split the field content of the theory
into physical and topological (non-propagating) sectors \cite{more}. In the $AdS$
background
fields $W_{i,i}$ and $B_{i, 1-i}$ are propagating, while $W_{i,
1-i}$ and $B_{i,i}$ on the contrary are non-dynamical
(topological). The fact that topological sector contains no
 propagating fields follows from the inspection of
the zero-form sector at free level. Indeed, within the unfolded
formulation physical degrees are stored in zero-forms. (For more
detail see e.g. \cite{Vasiliev:1999ba,RevD}). Upon resolution for
the dependence on the coordinates $Z^A$ the free level analysis
over $AdS$ background implies in the topological sector
\be
D_{\Omega}C_{i,i}(Y|x):=\dr C_{i,i}+[\Omega, C_{i,i}]_*=0\,,
\ee
which is the adjoint flatness condition. Here $\Omega$ is the
$AdS$-flat connection spanned by $Y$-bilinears
\be
\Omega=\ff i4\Omega_{AB}Y^{A}Y^{B}\,.
\ee
The adjoint covariant derivative acts on the finite-dimensional
modules, spanned by various homogeneous polynomials in $Y$, and
therefore the respective equations describe an infinite set of
topological systems, each describing at most a finite number of
degrees of freedom. In this respect, topological sector is
different from the dynamical sector in which zero-forms are valued
in the twisted-adjoint module, obeying the equations
\be\label{twad}
\dr C_{i,1-i}+\Omega*C_{i,1-i}-C_{i,1-i}*\pi(\Omega)=0\,,
\ee
where the antiautomorhism $\pi$ is defined to flip the sign of
undotted oscillators
\be
\pi(y,z,\bar y, \bar z)=(-y, -z, \bar y, \bar z)\,.
\ee
The twisted-adjoint module is infinite-dimensional being appropriate for the
description of relativistic fields carrying an infinite number of degrees of
freedom.

Note that the total numbers of degrees of freedom in the
topological and dynamical systems are the same, being described by
a pair of unrestricted functions of $Y^A$. This
is possible because the HS theory contains infinitely
many topological fields.

In principle, topological fields  can be truncated away from $4d$ HS theory
 by requiring $B_{ii}=W_{i, i+1}=S_{i,
i+1}=0$\footnote{Note that no such truncation is available in
$d=3$ theory of \cite{PV}.}. However, this leads to enormous
reduction of the space of HS theories since, as was recently
emphasized in \cite{meta}, the topological fields are in fact the
modules distinguishing between different theories. These fields
will also play a central role in this paper being associated with
chemical potentials in the partitions.

We apply  the bosonic truncation in this paper that allows one to
stay with a single copy of every integer spin in the spectrum.
This  is achieved by setting $k\bar k=1$ which is only consistent
if  $W$ and $B$ are bosons obeying $(W,B)(-y, \bar y, -z, \bar
z)=(W,B)(y, -\bar y, z, -\bar z)$ and $S(-y, \bar y, -z, \bar
z)=-S(y, -\bar y, z, -\bar z)$.

Apart from topological fields, the only freedom in
\eqref{HS1}-\eqref{HS5} is an arbitrary complex function
\be\label{func}
F_{*}(B)=\eta B+O(B^2)\,,
\ee
where $\eta$ is a constant phase factor, $|\eta|=1$. When all
higher powers of $B$ in \eqref{func} are discarded $\eta$ breaks
down parity of the theory unless $\eta=1$ or $\eta=i$ in which
cases one is left with the so-called A and B HS models
correspondingly \cite{SSAB}.

A convenient way to look at \eqref{HS1}-\eqref{HS5} that makes
issues of gauge invariance and consistency check straightforward
is the unification of all one-forms into a single field
\be
\W=\dr+W+S
\ee
which reduces \eqref{HS1}-\eqref{HS5} to
\begin{align}
&\W*\W=-i\theta_{\al}\wedge \theta^{\al}(1+F_*(B)*k\gk)-
i\bar\theta_{\dal}\wedge \bar\theta^{\dal}(1+\bar F_*(B)*\bar
k\bar
\gk)\,,\label{hs1}\\
&[\W, B]_*=0\,.\label{hs2}
\end{align}

As was shown recently in \cite{Vasiliev:1504}, written this way,
equations \eqref{hs1}-\eqref{hs2} admit straightforward
generalization allowing  to define the on-shell closed two-
and four-forms. The four-form is related presumably to on-shell HS
action thus being anticipated to play important role in the
$AdS/CFT$ analysis. The two-form ${\Lc}^2$, which is
of most interest in this paper, gives rise to a surface charge of a
particular solution of HS equations.
Namely, following \cite{Vasiliev:1504} we deform \eqref{hs1}-\eqref{hs2} as follows
\begin{align}
&\W*\W=-i\theta_{\al}\wedge \theta^{\al}(1+F_*(B)*k\gk)-
i\bar\theta_{\dal}\wedge \bar\theta^{\dal}(1+\bar F_*(B)*\bar
k\bar
\gk)+{\Lc}^2\,,\label{hsL1}\\
&\dr{\Lc}^2=0\,,\label{hsL2}\\
&[\W, B]_*=0\,,\label{hsL3}
\end{align}
where ${\Lc}^2(x)$ is a space-time two-form, independent of $Z$ and
$Y$ variables. According to  equations  \eqref{hsL1}, \eqref{hsL2},
${\Lc}^2$ is responsible for the following gauge transformations
\be\label{Lsym}
\gd{\Lc}^2(x)=\dr\gep(x)\,, \qquad \gd\W(Y,Z|x) =\gep(x)\,,
\ee
where gauge parameter $\gep(x)$ represents a space-time one-form.

Thus, by virtue of \eqref{hsL2} and \eqref{Lsym} ${\Lc}^2$ provides
a gauge-invariant surface charge via integration over a two-cycle $\Sigma$
\be\label{defcharge}
Q=\int_\Sigma {\Lc}^2\,.
\ee
Here ${\Lc}^2$ is expressed in terms of HS fields $W$ and $B$
through \eqref{hsL1}. By construction ${\Lc}^2$ is $Y$ and
$Z$ -- independent. At free level this implies that spin-1 field
only can contribute to the conserved charge, while all higher
spins affect through interaction. This fact reveals essential
difference between \eqref{defcharge} and standard canonical
constructions for gravity and HS fields based on the
action principle which  either reproduce asymptotic charge
(equivalently exact charge of any spin for a free theory) or imply
existence of a global symmetry. Let us stress that \eqref{defcharge}
is exactly conserved irrespectively of any leftover
global symmetry of a particular solution.

It might seem that the simplest free HS charges for Fronsdal
fields escape from \eqref{defcharge}. We will see that those can
be still obtained using  the topological sector of
\eqref{hsL1}-\eqref{hsL3}. In this case, the topological fields
play a role of a book-keeping device for contracting indices to
combine all closed forms into a single spin-one form that however
depends on the Killing tensors associated with the topological
fields whose field equations are just those of the Killing
tensors. In this respect it is not necessary to take into account
the back reaction of the the topological fields on the matter ones
to define asymptotic charges. Instead, as shown below, it is enough
to plug the Killing tensors into the formula for topological
fields which thus lead to asymptotic charges defined in the usual
HS theory free of the topological fields.

However, in order to use proposed construction for the globally
closed forms, the effect of back reaction of the propagating
fields on the topological ones should be fully taken into account.
This means that it is not clear how to define closed forms of
spins greater than one if the topological fields are not included
into the model. As speculated in the end of Section 5, this
phenomenon may have important implications for the resolution of
the information paradox.

Denoting $\go(y, \bar y |x)=W(Y, 0|x)$ and $C(y, \bar
y|x)=B(Y, 0|x)$, within perturbation theory around $AdS_4$
vacuum one arrives at the following schematic form of equations
resulting from \eqref{hsL1}, \eqref{hsL3}
\begin{align}
&\mathcal{R}(Y|x):=\dr\go+\go*\go={\Lc}^2-\Upsilon(\go, \go, C)-\Upsilon(\go, \go, C, C)-\dots\,,\label{dw}\\
&\dr C+[\go, C]_*=-\Upsilon(\go, C)-\Upsilon(\go, C,
C)-\dots\,,\label{dC}
\end{align}
where $\Upsilon$ denotes interaction vertices that can be found
order by order from the perturbative expansion. A powerful method
of extracting such vertices has been recently proposed in
\cite{perturb}.

The second term on the l.h.s. of \eqref{dw} is identically zero
at $Y=0$ because it can be represented as $tr(\go*\go)=0$ (recall that $\go$ is a one-form).
Indeed, the star product admits the  supertrace operation which in the
bosonic case has the property of usual trace
\be
tr f(Y):=f(0)\,,\qquad tr\,(f(Y)*g(Y))=tr\,(g(Y)*f(Y))\,.
\ee

So one has the following equation determining ${\Lc}^2$
\be\label{L_dw}
{\Lc}^2(x)=\dr\go(0|x)+\Big(\Upsilon(\go, \go, C)+\Upsilon(\go, \go, C,
C)+\dots\Big)_{Y=0}\,.
\ee

Next, using gauge symmetry \eqref{Lsym} one can impose the {\it{canonical gauge}}
$\go(0|x)=0$ \cite{Vasiliev:1504}. Then a two-form ${\Lc}^2$ can be expressed as
\be\label{L2}
{\Lc}^2(x)=\Big(\Upsilon(\go, \go, C)+\Upsilon(\go, \go, C,
C)+\dots\Big)_{Y=0}\,.
\ee

Being closed, ${\Lc}^2$ may or may not be exact. In the former case it
results in zero charge \eqref{defcharge}. This happens if \eqref{L_dw},
considered as an equation for $\go(0|x)$, admits solutions with ${\Lc}^2=0$
and  $\go(0|x)$ regular on the integration cycle $\Sigma$. On the contrary,
if $\go(0|x)$ at ${\Lc}^2=0$ gets singular on $\Sigma$, then ${\Lc}^2$
may be nontrivial.

This situation is analogous to Dirac monopole problem. There, due to
Dirac string, vector-potential is ill-defined on any surface surrounding
a monopole. However, magnetic field strength remains regular
and provides a conserved magnetic charge via integration over the surface.
Here we have ill-defined Abelian spin-1 potential $\go(0|x)$ and
well-defined 'field strength' $\Big(\Upsilon(\go, \go, C)+\dots\Big)_{Y=0}$.
Imposing the canonical gauge allows one to work entirely in terms of the latter,
though in general it is not obligatory: alternatively, as in the monopole
problem, one can use the fiber bundle picture, covering the spacetime with several
charts. Nontriviality of ${\Lc}^2$ then results from patching
conditions. Since this analysis is somewhat more complicated,
we prefer to work in the canonical gauge.

A related issue that we would like to emphasize is that it may be
tempting to solve eq. \eqref{HS1} in a pure gauge form suggesting
that any flat $W(Z,Y|x)$ is gauge equivalent to $W=0$. Again, this
immediately entails $Q=0$. This manipulation should not be
expected to be globally well-defined, however. This argument
resurrects an old problem of admissible gauge transformations and
classes of functions in general in HS theories, see \cite{PV,
VasLoc, Ponom, Tar} for a recent account.

The main new point stressed in this paper is that the construction
of the invariant functionals associated with ${\Lc}^2$ admits a
natural extension to various chemical potentials $\xi$ associated
with modules of topological fields in the HS theory introduced in
\cite{more}. The respective partition
\be
\label{part}
Z = \exp - \int_\Sigma {\Lc}^2(\xi)
\ee
turns out to be independent of the variations of the integration
two-cycle $\Sigma$. In the asymptotic limit where  $\Sigma$ tends
to infinity and the theory becomes nearly free, this construction
is argued to reproduce usual asymptotic charges in GR \cite{BMS1, BMS2, BMS3, Barnich, BB}, as well as
their HS extension. For surface charges in the HS context see
\cite{BG} and
 \cite{HPTT, BHPTT, Camp1, Camp2}.
While it may not be immediately straightforward
comparing our charge with the canonical asymptotic charges based
on the action principle and the metric-like approach, there is a
simple argument in favor of its equivalence. Our conclusions are
in full agreement with those of \cite{BB} since asymptotically
conserved charges are shown to be represented by on-shell closed
two-forms expressed in terms of dynamical fields of the theory
equivalent to the Fronsdal fields in the metric-like approach. The
complete set of these two-forms expressed via (generalized) Weyl
tensors is the same in the metric-like and frame-like formalism.
Hence they necessarily coincide.

\section{Charges and partitions in HS equations}

\subsection{Charges}
In Einstein case (Ricci=0) with some Killing one-form
$\xi=\xi_{\mu}dx^{\mu}$ there exists a conserved charge given by
the Komar integral \cite{Komar}
\be\label{komar}
Q=\int_{\Sigma} \mathcal{K}\,,
\ee
where $\mathcal{K}$ is a two-form defined via Killing one-form $\xi$
\be\label{komarK}
\mathcal{K}=\star \dr\xi\,,
\ee
which is closed on the equations of motion. For non-zero
cosmological constant $\Lambda\neq 0$ Komar expression
\eqref{komar} no longer works and one needs either subtracting
infinities due to the volume proportional to $\Lambda$ in
\eqref{komar} \cite{Magnon}, or use other refined methods, see
e.g. \cite{Kastor, Barnich, BB}. This way one normally produces BH
charges such as mass and angular momentum
using appropriate exact isometries. The resulting mass is of
course linear in BH mass parameter $m$.

 At the linearized level the analog of GR isometries in HS theory
are global symmetries, that is the redundant symmetries
$\gep(Y|x)$ of a vacuum solution such that
$\gd_{\gep}\go=\gd_{\gep}C=0$.
In \cite{Ashtekar}, the two-form that generates asymptotic
symmetries in GR was expressed in terms of the Weyl tensor contracted with vector
fields representing asymptotic symmetries.  HS theory admits a
natural generalization of this construction. Indeed, consider the following exact two-form
\be\label{Rxi}
\dr tr(\xi*\widehat{n}\go_1)\,.
\ee
Here $\xi(Y|x)$ is an $AdS$ Killing tensor, i.e.
\be\label{as}
D_{\Omega}\xi(Y|x)=0\,,\qquad D_{\Omega}=\dr+[\ff
i4\Omega_{AB}Y^{A}Y^{B}, \bullet]_*\,
\ee
and $\go_1(Y|x)$ is a linear fluctuation of the HS connection
$\go$, $\Omega_{AB}$ is the background $AdS_4$ connection (more
details in the following section) and the operator $\widehat{n}$
is defined via
\be\label{n}
\widehat{n}F(Y)=\begin{cases}
F(Y), & N_{y}>\bar{N}_{\bar{y}},\\
0, & N_{y}=\bar{N}_{\bar{y}},\\
-F(Y), & N_{y}<\bar{N}_{\bar{y}},
\end{cases}
\ee
where $N_{y}$ and $\bar{N}_{\bar{y}}$ count the powers of $y$ and
$\bar{y}$ respectively
\be\label{Ndef}
y^{\alpha}\dfrac{\partial}{\partial y^{\alpha}}F(Y)=N_{y}F(Y),
\qquad\bar{y}^{\dot{\alpha}}\dfrac{\partial}{\partial\bar{y}^{\dot{\alpha}}}F(Y)=\bar{N}_{\bar{y}}F(Y)\,.
\ee
Using \eqref{as}-\eqref{Ndef}, one finds that
\be\label{Rxi2}
\dr
tr(\xi*\widehat{n}\go_1)=tr\{\xi*e^{\alpha\dot{\beta}}\bar{y}_{\dot{\beta}}\dfrac{\partial}
{\partial
y^{\alpha}}(\widehat{c}_2+2\widehat{c}_1+\widehat{c}_0)\omega_{1}
-\xi*e^{\alpha\dot{\beta}}y_{\alpha}\dfrac{\partial}{\partial\bar{y}^{\dot{\beta}}}(\widehat{c}_{-2}
+2\widehat{c}_{-1}+\widehat{c}_0)\omega_{1}+\xi*\widehat{n}D_{\Omega}\omega_{1}\}\,,
\ee
where $\widehat{c}_n$ is the projector to the proper (sub)diagonal in the $(y,\bar{y})$-space
\be
\widehat{c}_n F(Y):=\begin{cases}
F(Y), & N_{y}-\bar{N}_{\bar{y}}=n,\\
0, & N_{y}-\bar{N}_{\bar{y}}\neq n\,.
\end{cases}
\ee
$\go_1$ satisfies what is known as the First on-mass-shell theorem
\be\label{OnshTh}
D_{\Omega}\go_1(y, \bar y|x)=-\ff{i\eta}{4}e_{\gga}{}^{\dal}\wedge
e^{\gga\dgb}\ff{\p^2}{\p\bar{y}^{\dal}\p\bar{y}^{\dgb}}C(0, \bar
y|x)-\ff{i\bar\eta}{4}e^{\al}{}_{\dgga}\wedge e^{\gb\dgga}
\ff{\p^2}{\p y^{\al}\p y^{\gb}}C(y, 0|x)\,,
\ee
following from linearization of the nonlinear HS equations. Each term on the r.h.s. of \eqref{OnshTh}
represents an independent $D_{\Omega}$-cohomology.

Components of the 1-form $\go_1(Y|x)$ with $N_{y}=\bar{N}_{\bar{y}}$ describe free bosonic Fronsdal fields
(in terms of base vector indices $\go_1^{\alpha(n),\dot{\alpha}(n)}(x)$ corresponds to double-traceless
rank-$(n+1)$ tensor $\phi^{\underline{a}(n+1)}(x)$), while components with  $N_{y}=\bar{N}_{\bar{y}}\pm1$
describe fermionic ones.
That is $\widehat{c}_0\go_1(Y|x)$
projects out bosonic Fronsdal fields. Similarly, by virtue of \eqref{OnshTh},
$\widehat{c}_{\pm 2n}\go_1$ extracts $n$-th derivatives thereof. For fermions
this is done by $\widehat{c}_{\pm 1}\go_1$ and $\widehat{c}_{\pm (2n+1)}\go_1$, respectively
(for more detail see \cite{Vasiliev:1999ba} and references therein).

Now we define closed non-exact two-form, corresponding to the last term
from  the r.h.s. of \eqref{Rxi2}
\be\label{Rxi3}
\mathcal{R}_{\xi}:=tr(\xi*\widehat{n}D_{\Omega}\omega_{1})\,,
\ee
which according to \eqref{OnshTh} equals
\be\label{Rxi4}
\mathcal{R}_{\xi}=tr \{ \xi*(\ff{i\eta}{4}e_{\gga}{}^{\dal}\wedge
e^{\gga\dgb}\ff{\p^2}{\p\bar{y}^{\dal}\p\bar{y}^{\dgb}}C(0, \bar
y|x)-\ff{i\bar\eta}{4}e^{\al}{}_{\dgga}\wedge e^{\gb\dgga}
\ff{\p^2}{\p y^{\al}\p y^{\gb}}C(y, 0|x)) \} \,.
\ee
That $\mathcal{R}_{\xi}$ is closed follows from the linearized equations
of motion on $C(Y|x)$ \eqref{dC}, which encode Bianchi identities and Maxwell equations.
$\mathcal{R}_{\xi}$ generates asymptotic charge via integration
over two-dimensional cycle $\Sigma^2$,
\be\label{acharge}
Q_{\xi}=\int_{\Sigma^2} \mathcal{R}_{\xi}\,.
\ee
Charge \eqref{acharge} provides a HS generalization of conserved
quantities in asymptotic $AdS$ defined in \cite{Ashtekar}. The
analogy is straightforward since charges of \cite{Ashtekar} arise
after integration of a two-form made of Weyl tensor and a symmetry
parameter. In our approach HS Weyl tensors naturally appear on the
r.h.s of \eqref{dw} in perturbative expansion around $AdS$
background as  explained in more detail in the next section.

Charge \eqref{acharge} is built of HS Weyl tensors $C(y, 0|x)$, $C (0,\bar y|x)$ which
correspond to $s$-th derivatives of spin-$s$ Fronsdal fields. At the same time there are
canonical one-derivative asymptotic Fronsdal charges, obtained in \cite{BG}. The relation
between two constructions is provided by Eq.~\eqref{Rxi2} implying that
\be
\label{rel}
\mathcal{R}_{\xi}=-tr\{\xi*(e^{\alpha\dot{\beta}}\bar{y}_{\dot{\beta}}\dfrac{\partial}
{\partial y^{\alpha}}(\widehat{c}_2+2\widehat{c}_1+\widehat{c}_0)
-e^{\alpha\dot{\beta}}y_{\alpha}\dfrac{\partial}{\partial\bar{y}^{\dot{\beta}}}(\widehat{c}_{-2}
+2\widehat{c}_{-1}+\widehat{c}_0))\omega_{1}\} +\dr
tr(\xi*\widehat{n}\go_1).
\ee
Since $\dr$-exact term does not contribute to the charge,
\eqref{acharge} can be equivalently expressed in terms of Fronsdal
fields and their first derivatives. This also explains a
peculiarity of the spin-1 sector. Spin-1 is described by
$Y$-independent connection $\go_1 (x)$, so it drops out from
\eqref{Rxi3}. However, nonzero spin-1 contribution can be embedded
into \eqref{Rxi4} with $C_{AB}Y^A Y^B (x)$. The reason is that
first derivatives of $\go_1(x)$ are already described by
gauge-invariant $C_{AB}Y^A Y^B (x)$, which is the Faraday tensor,
so there is no representation of spin-1 current in the form
\eqref{rel}, while \eqref{Rxi2} becomes an identity 0=0.

An asymptotically covariantly  constant parameter $\xi(Y|x)$,
$D\xi_{{z\to 0}}=0$ generates the asymptotically $\dr$-closed
${\mathcal{R}}_{\xi}$ and therefore asymptotically conserved
charge \eqref{acharge}. Here $\xi$ plays an auxiliary role of the
parameter that has nothing to do with a given HS solution. This
simple construction suggests an interesting possibility for the
study of BH charges in HS theory.

The appearance of the parameter $\xi$ that approaches $AdS$ (HS)
global symmetry parameter is natural in the HS system
\eqref{HS1}-\eqref{HS5} with the topological sector. The key
observation is that the form of the linearized equations in the
topological sector precisely coincides with that of the asymptotic
symmetry parameters in (\ref{as}).  This suggests an idea that the
topological fields can play a role of generalized chemical
potentials $\xi$ conjugated to HS charges.

At the nonlinear level the two sectors become entangled with
topological fields sourcing the  physical ones. The two-form
${\Lc}^2$ defined as \eqref{L2} at $Y=0$ is still closed in
presence of topological fields. This allows us to introduce an
invariant partition (\ref{part}) that depends on the modules of a
chosen solution like the BH mass as well as on the chemical
potentials $\xi$ identified with the topological fields. In this
setup, the partition is insensitive to local variations of the
integration cycle giving the same result upon integration at
infinity, where it should reduce to the asymptotic charges, and
over any other cycle in the same homotopy class including the horizon.

Naively, the existence of the closed form ${\Lc}^2$ in the
nonlinear theory and the surface-independent representation for
the partition (\ref{part}) may look surprising and even
impossible. However one should take into account the following two
facts. First of all, the functional ${\Lc}^2(\xi)$ is not a local
object in terms of dynamical and topological fields, containing an
infinite expansion in powers of derivatives of the fields in
$AdS_4$. Secondly, nonzero topological fields $\xi$ will affect
the form of the original (e.g. BH) solution at $\xi=0$. So, to
find a full partition (\ref{part}) for certain chemical potentials
$\xi$ one has to find the respective solution at $\xi\neq 0$.

In this paper we compute the vacuum value of ${\Lc}^2$ at $\xi=0$
to illustrate how the computation goes. It is simple to find the
contribution to  ${\Lc}^2(\xi)$ linear in $\xi$, associated in
particular with asymptotic charges and we present the explicit
formulae for that case as well. Computation of the nontrivial
charges to higher orders will be given elsewhere.

\subsection{Vacuum partition}

To extract ${\Lc}^2$ to the lowest order one truncates away the topological sector
and starts with the proper
vacuum solution of \eqref{HS1}-\eqref{HS5} that reads
\be
B^{(0)}=0\,,\quad S^{(0)}_A=Z_A\,,\quad W^{(0)}=\Omega=\ff
i4\Omega^{AB}Y_AY_B\,,
\ee
where the $AdS_4$ flat vacuum connection is given by
\be
\Omega=\ff{i}{4}(\go_{\al\gb}y^{\al}y^{\gb}+
\bar{\go}_{\dal\dgb}\bar{y}^{\dal}\bar{y}^{\dgb}+ 2\gl
e_{\al\dal}y^{\al}\bar{y}^{\dal})
\ee
that  gives as a consequence of  \eqref{HS1} equations for the
background $AdS_4$ space
\begin{align}
&de_{\al\dal}+\go_{\al}{}^{\gga}\wedge e_{\gga\dal}+
\bar{\go}_{\dal}{}^{\dgga}\wedge e_{\al\dgga}=0\,,\label{sT}\\
&d\go_{\al\gb}+\go_{\al}{}^{\gga}\wedge\go_{\gga\gb}+\gl^2
e_{\al}{}^{\dgga}\wedge e_{\gb\dgga}=0\label{sR}
\end{align}
and complex conjugate, where $\gl$ is the $AdS_4$ cosmological
parameter. The following free-level analysis is straightforward
(see e.g. \cite{Vasiliev:1999ba}) and results in
$B^{(1)}(Y,Z|x)=C(Y|x)$ that fulfils the twisted adjoint covariant
constancy condition (\ref{twad}).

As mentioned before, the HS potentials resided in
$\go(Y|x)=W^{(1)}(Y,Z|x)_{Z=0}$ obey First on-mass-shell Theorem
\be\label{comt}
D_{\Omega}\go(y, \bar y|x)=-\ff{i\eta}{4}e_{\gga}{}^{\dal}\wedge
e^{\gga\dgb}\ff{\p^2}{\p\bar{y}^{\dal}\p\bar{y}^{\dgb}}C(0, \bar
y|x)-\ff{i\bar\eta}{4}e^{\al}{}_{\dgga}\wedge e^{\gb\dgga}
\ff{\p^2}{\p y^{\al}\p y^{\gb}}C(y, 0|x)\,.
\ee
It is clear that the left-hand side of \eqref{comt} is a closed
two-form at $Y_A=0$. Therefore, the vacuum contribution to
${\Lc}^2$ at the free level is
\be\label{Lfree}
{\Lc}^2=-\ff{i}{4}\Big( \eta e_{\gga}{}^{\dal}\wedge
e^{\gga\dgb}\ff{\p^2}{\p\bar{y}^{\dal}\p\bar{y}^{\dgb}}C(0, \bar
y|x)+\bar\eta e^{\al}{}_{\dgga}\wedge e^{\gb\dgga} \ff{\p^2}{\p
y^{\al}\p y^{\gb}}C(y, 0|x)\Big)_{Y=0} \,.
\ee
The field $C(y, \bar y|x)$ contains matter fields and gauge-invariant
information of all free HS fields. The spin-$s$
contribution is captured by the following components
\be
\Big(y^{\al}\ff{\p}{\p y^{\al}}-\bar{y}^{\dal}\ff{\p}{\p
\bar{y}^{\dal}} \Big)C(y, \bar y|x)=\pm 2s C(y, \bar y|x)\,.
\ee
In other words, the components $C_{\al(2s+m), \dal(m)}$ and
$C_{\al(m), \dal(m+2s)}$ for any $m$ in
\[
C(y,\bar y|x)=\sum_{m,n=0}^{\infty}\ff{1}{m!
n!}C_{\al(m),\dal(n)}(y^{\al})^m(\bar{y}^{\dal})^n
\]
correspond to a spin-$s$ field and all its on-shell derivatives.
Particularly, purely (anti)holomorphic components stored in
$C(y,0|x)$ ($C(0,\bar y|x)$) are the so-called generalized HS Weyl
tensors $C_{\al(2s)}$ and $\bar{C}_{\dal(2s)}$. For example, for
$s=2$, $C_{\al_1\dots \al_4}$ and $\bar{C}_{\dal_1\dots \dal_4}$
are the spinor components of the usual gravity Weyl tensor.
Indeed, being totally symmetric the spin-tensors $C_{\al_1\dots
\al_4}$ and $\bar{C}_{\dal_1\dots \dal_4}$ are equivalent to
totally traceless window-like Young tensor $C_{ab,cd}$ in the
vector language.  For $s=1$, one has $C_{\al\gb}$ and
$\bar{C}_{\dal\dgb}$ being the spinor version of the antisymmetric
Maxwell tensor $C_{ab}=-C_{ba}$. For integer $s>2$,
both $C_{\al(2s)}$ and $\bar{C}_{\dal(2s)}$ encode totally
traceless two-row rectangular Young diagram in the vector language
describing the spin-$s$ Weyl tensors. These components are constrained by
the generalized Bianchi identities for any $s$ arising from
equation of motion \eqref{twad}. All the rest mixed components
$C_{\al(m), \dal(n)}$ can be expressed via the derivatives of HS
Weyl tensors through \eqref{twad}.

Therefore, the only nonzero contribution to ${\Lc}^2$ at the free
level comes from the Maxwell tensor $C_{\al\gb}$ and
$\bar{C}_{\dal\dgb}$ of the Abelian spin-one field
\be\label{L}
{\Lc}^2=-\ff{i\eta}{4} e_{\gga}{}^{\dal}\wedge e^{\gga\dgb} \bar
C_{\dal\dgb}-\ff{i\bar\eta}{4}e^{\al}{}_{\dgga}\wedge e^{\gb\dgga}
C_{\al\gb}\,.
\ee

The space-time contribution in \eqref{L} can be split into
local and nonlocal part. Indeed, slicing four-dimensional space
time into boundary coordinates $\overrightarrow{x}$ and bulk
direction $z$ as $x^{a}=(z, \overrightarrow{x})$, the two-form
$e\wedge e$ has the following schematic components $e_{z}\wedge
e_{\overrightarrow{x}}$ being nonlocal and
$e_{\overrightarrow{x}}\wedge e_{\overrightarrow{x}}$ being local
from the boundary perspective. More precisely, each contribution
can be easily singled out using Poincar$\mathrm{\acute{e}}$ coordinates
\be
ds^2=\ff{1}{z^2}(4\gl^2 dz^2+dx_{i}dx^{i}).
\ee
A convenient choice for the background fields is
\be\label{Poincare}
\go_{\al\gb}=\ff{i\gl}{2z}dx_{\al\gb}\,,\quad
\bar{\go}_{\dal\dgb}=-\ff{i\gl}{2z}dx_{\dal\dgb}\,,\quad
e_{\al\dgb}=\ff{1}{2z}(-dx_{\al\gb}\gd_{\dal}{}^{\gb}+i\gl^{-1}\gep_{\al\dgb}dz)\,,
\ee
where we have introduced the boundary coordinates
$x_{\al\gb}=x_{\gb\al}$ and the antisymmetric
$\gep_{\al\dal}=-\gep_{\dal\al}$ that explicitly breaks Lorentz
symmetry $X_{\al\dal}=(iz\gep_{\al\dal}, x_{\al\gb}\gd_{\dal}{}^{\gb})$.
Substituting the vierbein from \eqref{Poincare} into \eqref{L} one
finds
\be
{\Lc}^2={\Lc}^{2}_{loc}+{\Lc}^2_{nloc}\,,
\ee
where
\begin{align}
&{\Lc}^2_{loc}=-\ff{i}{16z^2}dx_{\gga}{}^{\al}\wedge
dx^{\gga\gb}(\eta\bar{C}_{\al\gb}+\bar\eta C_{\al\gb})\,,\label{loc}\\
&{\Lc}^2_{nloc}=\ff{\gl^{-1}}{8 z^2}dx^{\al\gb}\wedge dz (\eta
\bar{C}_{\al\gb}-\bar\eta C_{\al\gb})\,.\label{nloc}
\end{align}

Note that, as pointed out in \cite{Vasiliev:1504}, the nonlocal
contribution vanishes in particular in parity invariant $A$ model
with $\eta=1$. Indeed, in this case ${\Lc}^2_{nloc}$ remains
invariant under parity transform, while the right-hand side of
\eqref{nloc} changes its sign, therefore ${\Lc}^2_{nloc}=0$ for $
A$ model. A way out to have nonlocal contribution to the vacuum
charges in this important case is to keep the phase $\eta=\exp
i\phi$ arbitrary and then define
\be\label{Qnloc}
{L}^2_{nloc\, A}=\ff12\ff{\p {\Lc}^2}{\p\phi}_{|\phi=0}\,.
\ee
This procedure is to large extent analogous to considering a
variation over topological fields hence indicating that the full
partition function is likely to be nontrivial even in cases with
vanishing vacuum partitions. Note that the effect of this differentiation
is the same as of the insertion of the operator $\widehat{n}$ into (\ref{Rxi2}).

\subsection{Topological sector contribution}

Let us look at the effect that topological sector of HS equations
brings into the partition. The most general bosonic theory
is given by \eqref{hs1}-\eqref{hs2} with some arbitrary complex
function
\be\label{F}
F_*(B)=\eta B+\mu B*B+\dots\,,
\ee
where $\eta$ and $\mu$ are arbitrary complex parameters and  we assume
\be
B=B_{ph}(k+\bar k)+B_{top}(1+k\bar k)\,,\qquad W=W_{ph}(1+k\bar
k)+W_{top}(k+\bar k)\,.
\ee
In this paper we consider theories with essentially
nonlinear $F_*(B)$.
As  pointed out in \cite{Vasiliev:1504}, though terms nonlinear
in $B$  in \eqref{F} taken in the conformal frame may lead  to
boundary divergencies at conformal
infinity, this  does not imply any inconsistency of the theory in the bulk. Hence we do
not expect  it to affect the consideration of this paper. To fully clarify this issue it is
necessary to analyze in detail nonlinear corrections to
the conserved charge which analysis is beyond the scope of this
paper.

The closed two-form as defined in \eqref{L2}
\be\label{L2top}
{\Lc}^{2}=\ff{\p^2}{\p{k}\p\bar{k}}(W*W)_{(Y,\,Z\,=0)}
\ee
depends now not only on physical HS fields $\go(y, \bar y|x)$ and
$C(y, \bar y|x)$ but also on topological modules $\xi$ stored in
$C^{top}(y, \bar y|x)$. At the free level the topological sector
is decoupled and therefore does not contribute to \eqref{L2top}.
Indeed, linearized equations are
\begin{align}
&\dr C+\Omega*C-C*\pi(\Omega)=0\,,\\
&\dr C^{top}+[\Omega, C^{top}]_*=0\,.\label{Ctop}
\end{align}
Starting from second order the topological contribution is
nontrivial and intertwined with the HS one.

Terms bilinear in $C$
and $C^{top}$ appear in a simple way
\be\label{L2CC}
{\Lc}^2(\xi)=-\ff{i}{4}\left(\mu e_{\gga}{}^{\dal}\wedge
e^{\gga\dgb}\ff{\p^2}{\p\bar{y}^{\dal}\p\bar{y}^{\dgb}}\left(C^{top}(\xi)*C+
C*\pi(C^{top}(\xi))\right)+ c.c.\right)\,,
\ee
coming from the second-order piece in $F_{*}(B)$. More precisely,
taking $F_*(B)$ in the form \eqref{F} and recalling that $B\sim
C\oplus C^{top}$ one notes the First on-shell theorem \eqref{comt}
in this case accumulates the cross $C\times C^{top}$ piece
proportional to $\mu$. Eq. \eqref{L2CC} provides a perturbative
contribution to the conserved charge bilinear in $C$, $C^{top}$.
It therefore can be viewed as an asymptotic two-form generated by
Killing parameters $\xi$ stored in $C^{top}$ for HS theory without
topological backreaction. Indeed, what eq. \eqref{L2CC} says is
that for free (equivalently asymptotic) HS equations there is a
two-form which is $\dr$-closed on free HS equations.

Let us show now
 how one can reproduce this way an asymptotic charge for gravity
analogous to the one of \cite{Ashtekar}. In the construction of
\cite{Ashtekar} a closed $(d-2)$-form was built out of rescaled
Weyl tensor and a conformal Killing vector. For that let us choose
$C^{top}$ in the following form
\be\label{Ctop2}
C^{top}=\xi_{AB}Y^{A}Y^{B}\,,
\ee
where $D_{\Omega}\xi_{AB}=0$ so that $C^{top}$ verifies
\eqref{Ctop}. Though for gravity case the Weyl module should be
restricted to spin $s=2$ sector,  let us keep all spins excited
to see which of those will contribute to \eqref{L2CC}. Simple
star-product calculation of \eqref{L2CC} with $C^{top}$ from
\eqref{Ctop2} results in
\be
{\Lc}^{2}(\xi)={\Lc}^2_{s=0}(\xi)+{\Lc}^2_{s=2}(\xi)\,,
\ee
where
\begin{align}
&{\Lc}^2_{s=0}(\xi)=-i\mu e_{\gga}{}^{\dal}\wedge
e^{\gga\dal}\left(\bar{\xi}_{\dal\dal}C(x)-2i\xi^{\gb}{}_{\dal}C_{\gb\dal}(x)-\ff12\xi^{\gb\gb}
C_{\gb\gb, \dal\dal}(x)\right)+c.c.\,,\\
&{\Lc}^2_{s=2}(\xi)=-\ff{i\mu}{2} e_{\gga}{}^{\dal}\wedge
e^{\gga\dal}\bar{\xi}^{\dal\dal}\bar{C}_{\dal(4)}+c.c.\label{spin2}
\end{align}
We see that in this case the two-form picks up only spin-zero and
spin-two contributions. Spin one gets cancelled due to twisting in
\eqref{L2CC}. There are no HS contribution as well because the
generating parameter $\xi_{AB}$ carries only two indices. Had one
picked some higher-rank spin tensors $\xi_{A_1\dots A_{n}}$ there
would be nonzero HS contribution to two-form \eqref{L2CC}.
Expression \eqref{spin2} reproduces the result of \cite{Ashtekar}
with the difference that the Weyl tensor in \eqref{spin2} is free
of conformal scaling and a conformal Killing vector is replaced
with a Killing parameter $\xi_{\al\gb}$. Let us stress that our
construction is general allowing to reproduce higher-derivative
contribution for higher-rank parameters $\xi$ in a straightforward
and uniform fashion. Example of spin $s=4$ conserved charge
arises for parameter $C^{top}=\xi_{A_1\dots A_6}Y^{A_1}\dots
Y^{A_6}$ and can be straightforwardly calculated using
\eqref{L2CC}
\be
{\Lc}^2_{s=4}(\xi)=-\ff{i\mu}{2} e_{\gga}{}^{\dal}\wedge
e^{\gga\dal}\bar{\xi}^{\dal(6)}\bar{C}_{\dal(8)}+c.c.
\ee

One comment now is in order. It may look odd that to
reproduce charge for a free Fronsdal field one extends minimal HS
system by adding topological sector. In fact there are many ways
in obtaining asymptotic charges. One of the simplest is given by
tracing HS curvature with external parameters like in
\eqref{Rxi}, \eqref{acharge} which does not require any extension
of the HS equations. These external parameters can be made a part
of moduli space of the theory being  included via topological
extension which paves a way to chemical potentials. On a formal
side such an extension allows one to generate linearized charges
of a minimal theory  where  $\mu$ in
\eqref{F} plays a role of a coupling-like parameter which however
 does not have an independent meaning.
Namely, its product with the topological field has to be
identified with the generalized asymptotic Killing tensor that
appears in the usual approach to asymptotic symmetries, which can
be defined in the  minimal HS theory, in which case
 only  a linear response to the $\mu$-dependent terms is needed.
However, beyond the linearized approximation,
these terms allow us to make the Killing tensor-dependent two-form
closed everywhere which is indeed impossible in the genuine
minimal HS model.

 Let us stress again that system \eqref{hsL1}-\eqref{hsL3} admits
 the unique totally conserved charge \eqref{defcharge} rather
than  an asymptotic one.

Let us now proceed with the calculation of vacuum contribution to
conserved charge \eqref{defcharge}. We start with a HS BH solution
corresponding to a rotating source in the first nontrivial order.

\section{Black holes in GR and HS theory}
Remarkable property of any $4d$ Einstein BH including those with
nonzero cosmological constant $\Lambda$ is its 'linearized'
nature. Namely, the BH Weyl tensor satisfies linearized and full
nonlinear equations simultaneously. This is manifestated by the
famous Kerr-Schild decomposition
\be
g_{mn}=g_{0mn}+\ff{M}{U}k_{m}k_{n}\,,
\ee
where $g_{0mn}$ is the background $(A)dS$ metric, $M$  is the
parameter of mass, $U$ is some function and $k_m$ is the
Kerr-Schild vector that brings Einstein equations to the
linearized Fierz-Pauli.

At the level of curvature tensor such BHs and their Petrov D-type
analogs as well as the HS generalization originate from a single
$(A)dS$ global symmetry parameter \cite{DMV}. To recall the basic
construction consider $d$-dimensional $(A)dS$ space described by
the structure equations
\begin{align}
&dw_{ab}+w_{a}{}^{c}\wedge w_{cb}=\Lambda e_{a}\wedge e_{b}\,,\label{R}\\
&De_{a}=de_{a}+w_{a}{}^{b}\wedge e_{b}=0\,,\label{T}
\end{align}
where $w_{ab}=-w_{ba}$ and $e_{a}$ are one-forms of the Lorentz
connection and the vielbein, respectively. A nice feature of
Cartan form \eqref{R}-\eqref{T} for $(A)dS$ geometry is its
explicit local gauge invariance
\be
\gd w_{ab}=D\gk_{ab}+\Lambda(v_a e_b-v_b e_a)\,,\qquad \gd
e_{a}=Dv_a-\gk_{ab}e^b\,,
\ee
with arbitrary zero-forms $\gk_{ab}=-\gk_{ba}$ and $v_a$,
allowing one to fix $(A)dS$ global symmetries requiring $\gd
w_{ab}=\gd e_a=0$,
\begin{align}
&D v_a=\gk_{ab}e^{b}\,,\label{Kill}\\
&D\gk_{ab}=-\Lambda(v_a e_b-v_b e_a)\label{der}\,.
\end{align}
Here $v_a$ is just a Killing vector since from \eqref{Kill} it follows
that $D_{a}v_b+D_{b}v_a=0$, while $\gk_{ab}=-\gk_{ba}$ is its
covariant derivative. $\gk_{ab}$ and $v_a$ can be packed into
$(d+1)\times(d+1)$ - dimensional matrix, forming an element of
$o(d-1,2)$ for $\Lambda<0$ or $o(d,1)$ for $\Lambda>0$. This
matrix will be called the $(A)dS$ global symmetry parameter. It
turns out that system \eqref{Kill}-\eqref{der} generates exact
solutions to Einstein and HS equations in terms of $\gk_{ab}$ and
$v_a$ fields including all GR BHs. Now we consider the case of
$AdS_4$ in more detail following \cite{DMV}.

\subsection{Four dimensions}
For $AdS_4$, the well-known isomorphism $o(3,2)\sim sp(4)$ allows
us to use the two-component spinor language that facilitates
analysis a lot and suits the HS needs. Introducing
$\gk_{\al\gb}=\gk_{\gb\al}$,
$\bar{\gk}_{\dal\dgb}=\bar{\gk}_{\dgb\dal}$ and $v_{\al\dal}$ as
spinor counterparts for $\gk_{ab}$ and $v_a$ fields, the spinor
analog of \eqref{Kill}-\eqref{der} reads
\begin{align}
&D^{L}\gk_{\al\gb}=-\gl^2(e_{\al}{}^{\dgga}\wedge
v_{\gb\dgga}+e_{\gb}{}^{\dgga}\wedge v_{\al\dgga})\,,\label{Dk}\\
&D^{L}v_{\al\dal}=-e^{\gga}{}_{\dal}\gk_{\gga\al}-
e_{\al}{}^{\dgga}\bar{\gk}_{\dgga\dal}\,,\label{Dv}
\end{align}
where $\Lambda=-4\gl^2$ and $D^{L}A_{\al\dal}=\dr
A_{\al\dal}+\go_{\al}{}^{\gb}A_{\gb\dal}+
\bar{\go}_{\dal}{}^{\dgb}A_{\al\dgb}$. Introducing then
\be\label{K}
\Omega_{AB}=\begin{pmatrix} \go_{\al\gb} & \gl
e_{\al\dgb}\\
\gl e_{\gb\dal} & \bar{\go}_{\dal\dgb}\end{pmatrix}\,,\qquad
K_{AB}=\begin{pmatrix} \gk_{\al\gb} & \gl
v_{\al\dgb}\\
\gl v_{\gb\dal} & \bar{\gk}_{\dal\dgb}\end{pmatrix}
\ee
$A,B=1,\dots,4$ one rewrites \eqref{sT}-\eqref{sR} and
\eqref{Dk}-\eqref{Dv} as
\begin{align}
&\dr\Omega_{AB}+\Omega_{A}{}^{C}\wedge\Omega_{CB}=0\,,\label{flat}\\
&\dr K_{AB}+\Omega_{A}{}^{C}K_{CB}+\Omega_{B}{}^{C}K_{AC}=0.\label{glob}
\end{align}
Equation \eqref{flat} is the $AdS_4$ flatness condition and
$K_{AB}$ that satisfies \eqref{glob} is an $AdS_4$ global symmetry
parameter. In accordance with the fact that $rank(sp(4))=2$ there
are two $sp(4)$ invariants
\be\label{inv}
C_2=Tr K^2\,,\qquad C_4=4 Tr K^4\,.
\ee
An important observation \cite{DMV1} is that \eqref{Dk}-\eqref{Dv}
generates a tower of solutions for free massless equations of all
spins $s=0, 1, 2, \dots$, with the following generalized HS Weyl
tensors
\be\label{sol}
C_{\al(2s)}=\ff{m_s\gl^{-2s}(\gk_{\al\al})^s}{2^ss!q^{2s+1}}\,,\qquad
\bar{C}_{\dal(2s)}=\ff{\bar{m}_s\gl^{-2s}(\bar{\gk}_{\dal\dal})^s}{2^s
s!\bar{q}^{2s+1}}\,,
\ee
where
\be\label{scal}
q=\ff{1}{2\gl^2}\sqrt{-\ff{\gk_{\al\gb}\gk^{\al\gb}}{2}}\,,\qquad
\bar{q}=\ff{1}{2\gl^2}\sqrt{-\ff{\bar{\gk}_{\dal\dgb}\bar{\gk}^{\dal\dgb}}{2}}
\ee
and $m_s$ are arbitrary constants.

To show how solutions \eqref{sol} can be reproduced from
\eqref{twad}, it is useful to rewrite equation \eqref{glob} in
terms of star product as
\be
D_{\Omega}(K_{AB}(x)Y^AY^B)=0\,.
\ee
Since $D_{\Omega}$ is a first-order differential operator both in $x$ and in $Y$,
any function of $\xi=K_{AB}(x)Y^AY^B$ enjoys covariant
constancy condition as well
\be
D_{\Omega}f(\xi)=0\,.
\ee
The solution to twisted-adjoint covariant condition \eqref{twad}
can be obtained by Fourier transform \cite{DV}
\be\label{Fourier}
C(y, \bar y|x)=f(\xi)*2\pi\gd^2(y)\,.
\ee
Generally, \eqref{Fourier} does not meet the reality condition,
still it enables one to extract HS Weyl tensors that do satisfy
proper reality condition as
\be
C(y, 0|x)=(f(\xi)*2\pi\gd^2(y))_{\bar y=0}\,,\qquad C(0, \bar
y|x)=(f(\xi)*2\pi{\gd}^2(\bar y))_{y=0}\,.
\ee
Straightforward Gaussian integration
\be
F(y, \bar y)*2\pi\gd^2(y)=\int d^2 u F(u, \bar
y)e^{-iu_{\al}y^{\al}}
\ee
gives \eqref{sol}, where particular values of $m_{s}$ depend on
$f(\xi)$.

The solutions resulting from the construction are of generalized
Petrov type D in a sense that all the fields $s\geq 1$ are made of
two principal spinors which are the null directions of
$\gk_{\al\gb}$. The class of inequivalent solutions is represented
by the conjugacy classes with respect to $Sp(4)$ adjoint action
\be\label{conjug}
K_{AB}\sim (U^{-1}KU)_{AB}\,,
\ee
where $U_{A}{}^{B}\in Sp(4)$. Particularly, different values of
\eqref{inv} correspond to different $Sp(4)$ orbits. Note, however,
that by normalizing $K_{AB}$ one can always set one of the
invariants, say $C_2$ to $\pm   1$ or $0$. There are nine
conjugacy classes of $o(3,2)$ classified in e.g. \cite{o32}.
Finally, although not obvious (see \cite{DMV}) for $s=2$ a class
of thus obtained solutions is covered by Carter-Plebanski metric
which captures all type-D solutions of General Relativity
including all BHs except for accelerated metrics. We can therefore
refer to HS generalization as to HS Carter-Plebanski solutions. An
interesting parallelism can be drawn with the BTZ BH, where its
type is driven by a conjugacy class of $o(2,2)$ parameter
\cite{BTZH}. The analogy is not direct though, since $d=4$
solutions in question are not topological.

\subsection{Examples}
To be more specific let us review some examples  of
physical significance incuding the most symmetric BHs, the Kerr BH and the
Carter-Plebanski solution.

\subsubsection{Schwarzschild, planar and hyperbolic}
The examples below fall out from the general scheme of \cite{DMV}.
These were not analyzed explicitly in \cite{DMV}, so we do it
here. Consider the $AdS_4$ metric in the following coordinates
\be
ds^2=-f_\gep dt^2+f_{\gep}^{-1}dr^2-\Lambda r^2
d\Sigma_{\gep}^2\,,
\ee
where $\gep=\pm 1, 0$,
\be
f_{\gep}=\gep-r^2\Lambda
\ee
and
\be
d\Sigma^2_{\gep}=\left\{ \begin{array}{ll}
-\Lambda^{-1}(d\theta^2+\sin^2\theta
d\phi^2) & {\rm for}\; \gep=1\\
dx^2+dy^2 & {\rm for}\; \gep=0\\
-\Lambda^{-1}(d\psi^2+\sinh^2\psi d\theta^2) & {\rm for}\; \gep=-1
\end{array}\right..
\ee
Recall, that $\Lambda=-4\gl^2$. Each value of $\gep$ is designed
to reproduce the Schwarzschild ($\gep=1$), the planar ($\gep=0$)
and the hyperbolic ($\gep=-1$) BHs. The hyperbolic solution is
often referred to as topological due to possibility for
quotienting over a discrete subgroup of the hyperbolic horizon to
yield arbitrary genus Riemann surfaces \cite{Emparan}.

In each case we take the following Killing vector
\be\label{time}
v^{m}=(1,0,0,0)=\ff{\p}{\p t}
\ee
and the vierbein
\be
e^{0}=f^{1/2}_{\gep}dt\,,\quad e^{1}=f^{-1/2}_{\gep}dr\,,\quad
e^{2}=\left\{ \begin{array}{ll} \sqrt{-\Lambda^{-1}}d\theta & \; \gep=1\\
dx & \; \gep=0\\
\sqrt{-\Lambda^{-1}}d\psi & \; \gep=-1
\end{array}\right.\,,\quad e^{3}=\left\{ \begin{array}{ll} \sqrt{-\Lambda^{-1}}\sin\theta d\phi & \; \gep=1\\
dy &\; \gep=0\\
\sqrt{-\Lambda^{-1}}\sinh\psi d\theta & \; \gep=-1
\end{array}\right.\,,
\ee
which results in the following spinorial form for global symmetry
parameter components
\be
\gk_{\al\gb}=2\gl^2 r\begin{pmatrix}1 & 0\\
0 & -1 \end{pmatrix}\,,\qquad
v_{\al\dgb}=-f^{1/2}_{\gep}\begin{pmatrix}1 & 0\\
0 & 1 \end{pmatrix}\,.
\ee
The corresponding $K_{AB}$ from \eqref{K} has a distinguishing
$Sp(4)$ invariant property
\be\label{symBH}
K_{A}{}^{C}K_{C}{}^{B}=-\gl^2\gep\gd_{A}{}^{B}
\ee
that along with the choice of the Killing vector \eqref{time}
singles out the most symmetric BH solutions. The
eigenvalues are
\be\label{eigen}
K_{A}{}^{B}\xi_{B}=\tilde{\gl}\xi_{A}\,,\qquad \tilde{\gl}=\pm
i\gl\sqrt{\gep}\,.
\ee
The Weyl tensors are given by \eqref{sol} with
\be
q_{\gep}=\bar{q}_{\gep}=r\,.
\ee

Let us note that solutions of invariant condition \eqref{symBH}
correspond to most symmetric GR BHs. In other words, should
$K_{AB}$ be more generic, so that \eqref{symBH} no longer holds
the resulting BHs would be less symmetric. Clearly \eqref{symBH}
respects the $Sp(4)$ conjugacy transformation \eqref{conjug}.
Still it provides more than one inequivalent global symmetry
parameter \eqref{glob}. Expression \eqref{time} corresponds to a
Killing vector which is time-like, space-like or null and fixes
algebraical type of $K_{AB}$ unambiguously. Let us also note that
the centralizer of $K_{AB}$ in $sp(4)$ generates all isometries of
the corresponding BH solution.

\subsubsection{Kerr} To single out the  Kerr case we take the  $AdS_4$ Boyer-Lindquist metric suitably adopted
to account for rotation
\be\label{BL}
ds^2=-\ff{\Delta_r}{\rho^2}(dt-\ff{a}{\Xi}\sin^2\theta
d\phi)^2+\ff{\rho^2}{\Delta_r}dr^2+\ff{\rho^2}{\Delta_{\theta}}d\theta^2+
\ff{\sin^2\theta\Delta_{\theta}}{\rho^2}(a
dt-\ff{r^2+a^2}{\Xi}d\phi)^2\,,
\ee
where
\begin{align}
&\rho^2=r^2+a^2\cos^2\theta\,,\\
&\Delta_r=(r^2+a^2)(1-\Lambda r^2)\,,\\
&\Delta_{\theta}=1+\Lambda a^2\cos^2\theta\,,\\
&\Xi=1+\Lambda a^2\,.
\end{align}
A convenient form of the vierbein field reads
\begin{align}
&e^0=\ff{\sqrt{\Delta_r}}{\rho}(dt-\ff{a\sin^2\theta}{\Xi}d\phi)\,,\qquad
e^1=\ff{\rho}{\sqrt{\Delta_r}}dr\,,\\
&e^2=\ff{\rho}{\sqrt{\Delta_{\theta}}}d\theta\,,\qquad
e^3=\ff{\sqrt{\Delta_{\theta}}\sin\theta}{\rho}(a\,
dt-\ff{r^2+a^2}{\Xi}d\phi)\,.
\end{align}
Taking the following Killing vector
\be\label{time2}
v^{m}=\ff{\p}{\p t}=(1, 0, 0, 0)
\ee
one finds the $sp(4)$ global symmetry parameter components to be
\be\label{KerrK}
\gk_{\al\gb}=2\gl^2(r-ia\cos\theta)\begin{pmatrix} 1 & 0\\
0 & -1\end{pmatrix}, v_{\al\dgb}=\ff{1}{\rho}\begin{pmatrix}
-\sqrt{\Delta_r}+a\sqrt{\Delta_{\theta}}\sin\theta & 0\\
0 &
-\sqrt{\Delta_r}-a\sqrt{\Delta_{\theta}}\sin\theta\end{pmatrix}\,.
\ee
Note now that Killing vector \eqref{time2} is time-like in
$AdS$ case and not sign-definite in $dS$. Indeed, its norm is
given by
\be
v_t\cdot v_t=-1+\Lambda(r^2+a^2 \sin^2\theta)\,,
\ee
which is sign-definite at $\Lambda\leq 0$. The $sp(4)$ invariants
\eqref{inv} and the eigenvalues \eqref{eigen} are
\begin{align}
&C_2=\Lambda (1-a^2\Lambda)\,,\qquad C_4-C_2^2=-4a^2\Lambda^3\,,\label{Kerr}\\
&\tilde{\gl}_{1,2}=\pm i(1+2a\gl)\,,\qquad \tilde{\gl}_{3,4}=\pm
i(1-2a\gl)\,.
\end{align}
Again, for $s=2$ we reproduce Kerr BH Weyl tensor in
\eqref{sol} with
\be\label{Kerrq}
q=r-ia\cos\theta\,,\qquad \bar q=r+ia\cos\theta\,.
\ee
In the Kerr case eq. \eqref{symBH} no longer holds and the
solution in $AdS$ is fixed by generic time-like symmetry
parameter \eqref{time2}. The global symmetry invariants
\eqref{Kerr} are expressed in terms of the  rotation parameter $a$.

\subsubsection{Carter-Plebanski}
To describe the most general case that captures all above cases as
different limits it is convenient to introduce two-parameter
$AdS_4$ metric in the Carter-Plebanski form, \cite{DMV}
\be\label{Carter}
ds^2=-\ff{\Delta_r}{r^2+y^2}(d\tau+y^2d\psi)^2+\ff{\Delta_y}{r^2+y^2}(d\tau-r^2d\psi)^2
+\ff{r^2+y^2}{\Delta_r}dr^2+\ff{r^2+y^2}{\Delta_y}dy^2\,,
\ee
where
\be
\Delta_r=r^2(-\Lambda r^2+\gep)+a^2\,,\quad \Delta_y=y^2(-\Lambda
y^2-\gep)+a^2\,.
\ee
From that metric the Carter-Plebanski solution is generated via
$\ff{\p}{\p\tau}$ Killing vector for which the norm equals
\be
v_{\tau}\cdot v_{\tau}=-\gep+\Lambda (r^2-y^2)\,.
\ee
Choosing the vierbein as
\be
e^0=\sqrt{\ff{\Delta_r}{r^2+y^2}}(d\tau+y^2 d\psi)\,,\quad
e^1=\sqrt{\ff{\Delta_y}{r^2+y^2}}(d\tau-r^2d\psi)\,,\quad
e^2=\sqrt{\ff{r^2+y^2}{\Delta_r}}dr\,,\quad
e^3=\sqrt{\ff{r^2+y^2}{\Delta_y}}dy
\ee
we find the generating $AdS_4$ global symmetry parameter in the
form
\be
\gk_{\al\gb}=2\gl^2\begin{pmatrix} y-ir & 0\\
0 & y-ir\end{pmatrix}\,,\qquad
v_{\al\dgb}=\ff{1}{\sqrt{r^2+y^2}}\begin{pmatrix} \sqrt{\Delta_r}
& \sqrt{\Delta_y}\\
\sqrt{\Delta_y} & \sqrt{\Delta_r}\end{pmatrix}
\ee
that corresponds to
\begin{align}
&C_2=\Lambda\gep\,,\qquad C_4-C_2^2=-4a^2\Lambda^3\,,\\
&\tilde{\gl}_{1,2}=\pm i\gl\sqrt{\gep+4a\gl}\,,\qquad
\tilde{\gl}_{3,4}=\pm i\gl\sqrt{\gep-4a\gl}\,.
\end{align}
and
\be
q=y-ir\,,\qquad \bar q=y+ir\,.
\ee
Note that Kerr solution is reproduced at
$\gep=1-\Lambda a^2$.

\subsection{Charges of Kerr BH}

Now we are in a position to obtain explicit expressions for conserved charges
\eqref{defcharge} of Kerr-like HS BH.

First, let us consider a vacuum contribution at the free level.
It is generated by \eqref{L} for Kerr-type Maxwell tensor \eqref{sol}
\be
C_{\al\gb}=\ff{m_1\gk_{\al\gb}}{2\gl^2 q^3}\,,
\ee
where $\gk_{\al\gb}$ and $q$ were calculated in \eqref{KerrK} and
\eqref{Kerrq}. Straightforward calculation puts \eqref{L} in the
following simple form
\be\label{L2Kerr}
{\Lc}^2=-i\Big(\ff{m_1\bar\eta}{q^2}+\ff{\bar{m}_1\eta}{\bar{q}^2}\Big)e^0\wedge
e^1+\Big(\ff{m_1\bar\eta}{q^2}-\ff{\bar{m}_1\eta}{\bar{q}^2}\Big)e^3\wedge
e^2\,.
\ee

Corresponding charge \eqref{defcharge} can be easily evaluated via
integration of
\eqref{L2Kerr} around $t=const$, $r\rightarrow\infty$ and is eventually
given by
\be
Q=4\pi\ff{m_1\bar{\eta}-\bar{m}_1\eta}{1+\Lambda a^2}\,.
\ee
As anticipated, the so obtained vacuum charge is zero for
$\eta=\bar\eta$ for the Kerr-case with $m_1=\bar{m}_1$. However it
is non-zero for parity violating theories. Let us also note, that
in this case  $Q$ agrees up to a numerical normalization with the
standard $s=1$ charge for $AdS$-Kerr-Newman BH, see e.g.
\cite{Klemm}. In the parity preserving case one uses
\eqref{Qnloc} to reproduce the charge.

Next we compute the lowest nontrivial topological contribution to the spin-2 charge,
provided by \eqref{spin2}. To this end we identify Killing parameter $\xi_{\alpha\beta}$
in \eqref{spin2} with $\varkappa_{\alpha\beta}$ from \eqref{KerrK} and
substitute there $s=2$ Weyl tensor \eqref{sol} built of the same $\varkappa_{\alpha\beta}$
\be
C_{\alpha\beta\gamma\delta}=\dfrac{m_{2}}{8\lambda^{4}q^{5}}
\varkappa_{(\alpha\beta}\varkappa_{\gamma\delta)}\,.
\ee
This yields
\be
{\Lc}_{s=2}^{2}=3i\lambda^{2}\left(\dfrac{m_{2}\bar{\mu}}{q^{2}}+
\dfrac{\bar{m}_{2}\mu}{\bar{q}^{2}}\right)e^{0}\wedge e^{1}-3\lambda^{2}
\left(\dfrac{m_{2}\bar{\mu}}{q^{2}}-\dfrac{\bar{m}_{2}\mu}{\bar{q}^{2}}
\right)e^{3}\wedge e^{2}\,.
\ee
Once again, to obtain a charge we integrate this around $t=const$, $r\rightarrow\infty$,
arriving at
\be
Q_{s=2}=3\pi\Lambda\dfrac{m_{2}\bar{\mu}-\bar{m}_{2}\mu}{1+\Lambda a^{2}}\,.
\ee
Now we can adjust the parameters $\mu$ and $\bar \mu$ to reproduce the proper
BH charge.

\section{Conclusion}

In this paper we conjecture that the partition associated with
solutions of HS theory is determined by the closed two-form
associated with the full HS theory that contains the topological
fields which from the thermodynamical perspective represent
chemical potentials conjugated to various higher-spin and
lower-spin charges. This construction not only properly reproduces
the asymptotic charges but also allows a proper deformation into
the bulk.

Specifically, in this paper we compute the vacuum contributions to
the partition at zero values of the chemical potentials at free
level and also the first order contribution of the chemical
potentials which allows us to extract at lowest order the BH
conserved charge for the proper choice of moduli parameters.

It is demonstrated that the invariant functional two-form defined
on-shell provides nontrivial vacuum partition. The reason why it
is not trivial is cohomological. Whenever solutions are globally
well-defined the corresponding functional is trivial. However, in
case of global obstruction it may be non-zero. That this is indeed
the case was illustrated with an example of the HS analog of
$AdS_4$ Kerr BH at the  linearized level. The corresponding global
charge turned out to be nonzero and agreed with the asymptotic ADM
behaviour. Let us stress that those solutions that are not
everywhere well-defined are often of most physical interest including
BHs, Coulomb potential, Dirac string and its gravity cousin, the
NUT solution.

We have also analyzed leading contribution of the chemical
potentials to asymptotic charge which are the parameters stored in
the topological sector of the theory. Their effect can be easily
captured to the first order in physical fields and gives a simple
formula that reproduces known result \cite{Ashtekar} for the case
of gravity. Yet the obtained result is uniformly applicable to the
asymptotic charge of any spin associated with an arbitrary rank
parameter. Therefore we propose an efficient tool for extracting
HS asymptotic charges. It would be interesting to study how
the proposed approach can be applied to other Lagrangian
HS systems like those of \cite{BSaction}.

We believe that the proposed approach will make it possible not
only evaluating nontrivial charges for the HS BH solutions of
\cite{DV}, \cite{IS} starting from the linear response to the
various chemical potentials but eventually can shed light on the
deep issue of the BH information paradox. The most important
ingredient of the developed formalism is that the charges are
independent of the integration cycle and hence may be integrated
equally well both at infinity and at the horizon. Particularly
this paves a way to HS analogs of thermodynamical Smarr formulas.
Another possibly relevant issue is that the physical
interpretation of various fields in the theory such as dynamical
and topological fields may depend on the behaviour of the vacuum
solution that can change the interpretation of the fields in
different regions of space-time. All this makes HS theory
extremely interesting for the study of BH physics. We hope to
discuss these intriguing issues in the future.

As a byproduct of our construction we obtained an explicit formula (\ref{rel}) relating
asymptotic charges expressed in terms of the generalized HS Weyl tensor with those
expressed in terms of Fronsdal fields.

\section*{Acknowledgements}

This research was supported in part by the RFBR Grant No 14-02-01172.
One of us (MV) is grateful to Per Sundell and especially to Glenn Barnich for
stimulating discussions of asymptotic charges during the School and Workshop
on Higher Spins, Strings and Dualities in
Quintay, Chili, December, 2015. We are also grateful to the referee for the suggestion
to work out an explicit relation between the two forms of asymptotic HS charges.


\begin{thebibliography}{10}
\bibitem{Vasiliev:1999ba} M.~A.~Vasiliev, \textit{Higher spin gauge theories: Star-product and AdS space},
[hep-th/9910096].

\bibitem{RevD} X. Bekaert, S. Cnockaert, Carlo Iazeolla, M.A.
Vasiliev, \textit{Nonlinear higher spin theories in various
dimensions}, [hep-th/0503128].

\bibitem{GYRev} S. Giombi, X. Yin, \textit{The Higher Spin/Vector Model
Duality}, J.Phys. A46 (2013) 214003, [arXiv:1208.4036].

\bibitem{KP} I. Klebanov, A. Polyakov, \textit{AdS dual of the critical O(N) vector
model}, Phys.Lett. B550 (2002) 213-219, [hep-th/0210114].

\bibitem{SS} E. Sezgin, P. Sundell, \textit{Holography in 4D (super) higher spin theories and a test via cubic scalar couplings
}, JHEP 0507 (2005) 044, [hep-th/0305040].

\bibitem{fronsdal} C. Fronsdal, \textit{Massless Fields with Integer Spin}, Phys.Rev. D18 (1978) 3624 UCLA/78/TEP/5.

\bibitem{BSaction} N. Boulanger, P. Sundell, \textit{An action principle
for Vasiliev's four-dimensional higher-spin gravity}, J.Phys. A44
(2011) 495402, [arXiv:1102.2219].

\bibitem{GY1} S. Giombi, X. Yin, \textit{Higher Spin Gauge Theory and Holography: The Three-Point
Functions}, JHEP 1009 (2010) 115, [arXiv:0912.3462].

\bibitem{GY2}  S. Giombi, X. Yin, \textit{ Higher Spins in AdS and Twistorial
Holography}, JHEP 1104 (2011) 086, [arXiv:1004.3736].

\bibitem{MZh1} J. Maldacena, A. Zhiboedov, \textit{Constraining Conformal Field Theories with A Higher Spin
Symmetry}, J.Phys. A46 (2013) 214011, [arXiv:1112.1016].

\bibitem{MZh2} J. Maldacena, A. Zhiboedov, \textit{Constraining conformal field theories with a slightly broken higher spin
symmetry}, Class.Quant.Grav. 30 (2013) 104003, [arXiv:1204.3882].

\bibitem{Vasiliev:1504} M. A. Vasiliev, \textit{Invariant Functionals
in Higher-Spin Theory}, [arXiv:1504.07289].

\bibitem{more} M.~A.~Vasiliev, \textit{More on equations of motion for interacting massless fields of all spins in 3+1 dimensions},
Phys. Lett. B 285 (1992) 225.

\bibitem{Ashtekar} A. Ashtekar, S. Das, \textit{Asymptotically Anti-de Sitter Space-times: Conserved
Quantities}, Class.Quant.Grav.17:L17-L30,2000, [hep-th/9911230].

\bibitem{DMV} V.E. Didenko, A.S. Matveev, M.A. Vasiliev, \textit{ Unfolded Dynamics and Parameter Flow of Generic AdS(4) Black
Hole}, [arXiv:0901.2172].

\bibitem{PV} S.F. Prokushkin, M.A. Vasiliev, \textit{Higher spin gauge interactions for massive matter fields in 3-D AdS
space-time}, Nucl.Phys. B545 (1999) 385, [hep-th/9806236].

\bibitem{meta} M.A. Vasiliev, \textit{Higher-Spin Theory and Space-Time
Metamorphoses}, Lect.Notes Phys. 892 (2015) 227-264,
[arXiv:1404.1948].

\bibitem{SSAB} E. Sezgin, P. Sundell, \textit{ Holography in 4D (super)
higher spin theories and a test via cubic scalar couplings}, JHEP
0507 (2005) 044,[hep-th/0305040].

\bibitem{perturb} V.E. Didenko, N.G. Misuna, M.A. Vasiliev,
\textit{Perturbative analysis in higher-spin theories},
[arXiv:1512.04405].

\bibitem{VasLoc} M.A. Vasiliev, \textit{Star-Product Functions in Higher-Spin Theory and
Locality}, JHEP 1506 (2015) 031, [arXiv:1502.02271].

\bibitem{Ponom} X. Bekaert, J. Erdmenger, D. Ponomarev and C. Sleight, \textit{Quartic AdS Interactions in Higher-Spin Gravity from Conformal Field Theory},
JHEP 1511 (2015) 149 [arXiv:1508.04292 [hep-th]].

\bibitem{Tar} M. Taronna, E.D. Skvortsov, \textit{On Locality, Holography and
Unfolding}, JHEP 1511 (2015) 044, [arXiv:1508.04764].

\bibitem{BMS1} H. Bondi, M.G.J. van der Burg, A.W.K. Metzner, \textit{Gravitational waves in general relativity. 7. Waves from axisymmetric isolated
systems}, Proc.Roy.Soc.Lond. A269 (1962) 21-52.

\bibitem{BMS2} R.K. Sachs, \textit{Gravitational waves in general relativity. 8. Waves in asymptotically flat space-times
}, Proc.Roy.Soc.Lond. A270 (1962) 103-126.

\bibitem{BMS3} R. Sachs, \textit{Asymptotic symmetries in gravitational
theory}, Phys.Rev. 128 (1962) 2851-2864.

\bibitem{Barnich} G. Barnich, S. Leclercq, P. Spindel,
\textit{Classification of surface charges for a spin two field on
a curved background solution }, Lett.Math.Phys. 68 (2004) 175-181,
[gr-qc/0404006].

\bibitem{BB} G. Barnich, F. Brandt \textit{Covariant theory of
asymptotic symmetries, conservation laws and central charges},
[hep-th/0111246].

\bibitem{BG} G. Barnich, N. Bouatta, M. Grigoriev, \textit{Surface charges and dynamical Killing tensors for higher spin gauge fields in constant curvature
spaces}, JHEP 0510 (2005) 010, [hep-th/0507138].

\bibitem{HPTT} M. Henneaux, A. Perez, D. Tempo, R. Troncoso, \textit{Chemical potentials in three-dimensional higher spin anti-de Sitter gravity},
JHEP 1312 (2013) 048, [arXiv:1309.4362].

\bibitem{BHPTT} C. Bunster, M. Henneaux, A. Perez, D. Tempo, R. Troncoso, \textit{Generalized Black Holes in Three-dimensional Spacetime},
JHEP 1405 (2014) 031, [arXiv:1404.3305].

\bibitem{Camp1} A. Campoleoni, M. Henneaux, \textit{Asymptotic symmetries of three-dimensional higher-spin gravity: the metric approach},
JHEP 1503 (2015) 143, [arXiv:1412.6774].

\bibitem{Camp2} A. Campoleoni, H.A. Gonzalez, B. Oblak, M. Riegler, \textit{Rotating Higher Spin Partition Functions and Extended BMS Symmetries},
JHEP 1604 (2016) 034, [arXiv:1512.03353].

\bibitem{Komar} A. Komar, \textit{Covariant Conservation Laws in General
Relativity}, Phys. Rev. 113, 934, 1959.

\bibitem{Magnon} A. Magnon, \textit{On Komar integrals in asymptotically anti-
de Sitter space-times}, J. Math. Phys. 26 , 3112 (1985).

\bibitem{Kastor} D. Kastor, \textit{Komar Integrals in Higher (and Lower)
Derivative Gravity}, Class.Quant.Grav.25:175007,2008,
[arXiv:0804.1832].

\bibitem{DMV1} V.E. Didenko, A.S. Matveev, M.A. Vasiliev,
\textit{Unfolded Description of AdS(4) Kerr Black Hole},
Phys.Lett. B665 (2008) 284-293.

\bibitem{DV} V.E. Didenko, M.A. Vasiliev, \textit{ Static BPS black hole in 4d higher-spin gauge theory
} Phys.Lett. B682 (2009) 305-315 Phys.Lett. B722 (2013) 389,
[arXiv:0906.3898].

\bibitem{o32} S. Holst, P. Peldan, \textit{Black Holes and Causal Structure in Anti-de
Sitter Isometric Spacetimes}, Class.Quant.Grav. 14 (1997)
3433-3452, [gr-qc/9705067].

\bibitem{BTZH} M. Banados, M. Henneaux, C. Teitelboim, J. Zanelli,
\textit{Geometry of the (2+1) black hole}, Phys.Rev. D48 (1993)
1506-1525, Phys.Rev. D88 (2013) 069902, [gr-qc/9302012].

\bibitem{Emparan} R. Emparan, \textit{AdS/CFT Duals of Topological Black Holes and the Entropy of Zero-Energy
States}, JHEP 9906 (1999) 036, [hep-th/9906040].

\bibitem{Klemm} M.M. Caldarelli, G. Cognola, D. Klemm, \textit{Thermodynamics of Kerr-Newman-AdS Black Holes and Conformal Field Theories},
Class.Quant.Grav. 17 (2000) 399-420, [hep-th/9908022].

\bibitem{IS} C. Iazeolla, P. Sundell, \textit{ Families of exact solutions to Vasiliev's 4D equations
with spherical, cylindrical and biaxial symmetry }, JHEP 1112
(2011) 084, [arXiv:1107.1217].



\end{thebibliography}
\end{document}